\documentclass[twocolumn,showpacs,preprintnumbers,amsmath,amssymb]{revtex4-1}

\usepackage{graphicx}
\usepackage{dcolumn}
\usepackage{bm}
\def\3he{$^3$He}
\def\4he{$^4$He}

\begin{document}

\preprint{}

\title{Dissipative superfluid mass flux through solid $^4$He}

\author{Ye. Vekhov}
\author{R.B. Hallock}%
\affiliation{%
Laboratory for Low Temperature Physics, Department of Physics,\\
University of Massachusetts, Amherst, MA 01003.
}%

\date{\today}

\begin{abstract}
The thermo-mechanical effect in superfluid helium is used to create an initial chemical potential difference, $\Delta \mu_0$, across a solid $^4$He sample. This $\Delta \mu_0$ causes a flow of helium atoms from one reservoir filled with superfluid helium, through a sample cell filled with solid helium, to another superfluid-filled reservoir until chemical potential equilibrium is restored.  The solid helium sample is separated from each of the reservoirs by Vycor rods that allow only the superfluid component to flow. With an improved technique, measurements of the flow, $F$, at several fixed solid helium temperatures, $T$, have been made as function of  $\Delta \mu$ in the pressure range 25.5 - 26.1 bar.  And, measurements of $F$  have been made as a function of temperature in the range $180 < T < 545$~mK for several fixed values of $\Delta \mu$. The temperature dependence of the flow above $100$~mK shows a reduction of the flux with an increase in temperature that is well described by $F = F_0^*[1 - a\exp(-E/T)]$.  The non-linear functional dependence $F \sim (\Delta \mu)^b$, with $b < 0.5$ independent of temperature but dependent on pressure, documents in some detail the dissipative nature of the flow and suggests that this system demonstrates Luttinger liquid-like one-dimensional behavior. The mechanism that causes this flow behavior is not certain, but is consistent with superflow on the cores of edge dislocations.
\end{abstract}

\pacs{67.80.-s, 67.80.B-, 67.80.bd, 71.10.Pm}
\maketitle

\section{Introduction}
\label{introduction}

The torsional oscillator measurements of Kim and Chan\cite{Kim2004a,Kim2004b} and their interpretation of the data to suggest the possible existence of the theoretically-predicted\cite{Penrose1956,Andreev1969,Chester1970,Guyer1971,Meisel1992} supersolid\cite{Leggett1970}, stimulated a considerable renewal of interest in the properties and behavior of solid \4he.  The supersolid interpretation has been questioned by a number of workers who suggest that some experiments carried out to date may show no clear or at most only weak  evidence for supersolid behavior\cite{Syshchenko2010,Reppy2010}.  Importantly, recent work by Chan's group with a Vycor cell coated with epoxy (instead of being enclosed in a container with small amounts of bulk solid present)\cite{Kim2012} and with cells designed to minimize the shear modulus effect\cite{Kim2014} has shown that the original interpretation of the solid helium in Vycor work was likely premature.  Rather than the observation of supersolid behavior, it is now believed that that the original Kim and Chan\cite{Kim2004a,Kim2004b} observations resulted from changes in the stiffness\cite{Day2007} of the bulk helium in the sample cell and the influence of this temperature-dependent stiffness on the torsional oscillator\cite{Maris2011,Beamish2012} itself.

Experiments designed to directly create flow in solid \4he in confined geometries by squeezing the solid lattice directly have not been successful\cite{Greywall1977,Day2005,Day2006,Rittner2009}.  We took a different approach and by the creation of chemical potential differences across bulk solid \4he samples in contact with superfluid helium we have demonstrated mass transport by measuring the mass flux, $F$, through a cell filled with solid \4he\cite{Ray2008a,Ray2009b}. We found evidence for flux at temperatures that extend to values above those where torsional oscillator or other experiments have focused attention.  These experiments,  for \4he with a presumed nominal 300 ppb \3he impurity content, revealed a dramatic collapse of  the flux\cite{Ray2010c,Ray2011a} on cooling through the vicinity of 75-80~mK, with evidence for some recovery at lower temperatures.

Although a brief report that covers a portion of the content we report here has appeared\cite{Vekhov2012}, here we describe in some detail our experiments and evolving understanding of the behavior of $F$ for $T \gtrsim$ 180 mK.  We apply a temperature difference, $\Delta T$, to create an initial chemical potential difference, $\Delta \mu_0$, between two superfluid-filled reservoirs in series with a cell filled with solid \4he. We then measure the behavior of the \4he flux through the solid-filled cell. This flux results from the imposed $\Delta T$ and changes with time as the pressure difference between the two reservoirs, $\Delta P$,
changes (due to the fountain effect)
and the chemical potential difference between the two reservoirs, $\Delta \mu$,
\begin{equation}
\Delta \mu = m_4[\int(dP/\rho) - \int(sdT)],
\label{Eq_chemical}
\end{equation} 
changes from the initially imposed $\Delta \mu_0$ to zero.
Here  $m_4$ is the \4he mass, $\rho$ is the density and $s$ is the entropy per unit mass.  
We find that the flux, $F$, is not independent of $\Delta \mu$, but can be described at fixed solid \4he temperature by $F = A(\Delta \mu)^b$, where $A$ and $b$ are fitting parameters and where $A$ is found to be a decreasing function of increasing temperature and $b$ is temperature independent, but both $A$ and $b$ depend on pressure. This non-linear behavior of $F$ as function of $\Delta\mu$ above $T \sim $100 mK provides evidence that  the flux in the \4he solid may be due to a conduction process that arises as a result of the presence of bosonic Luttinger liquid behavior.  The precise nature of what actually carries the flux remains uncertain, and liquid channels have been proposed as a possibility\cite{Sasaki2008},  but the results to date are consistent with the flux being conducted by the cores of edge dislocations\cite{Soyler2009}.

\section{Experimental technique}
\label{technique}

\begin{figure}[htb]
\includegraphics[width=0.6\linewidth,keepaspectratio]{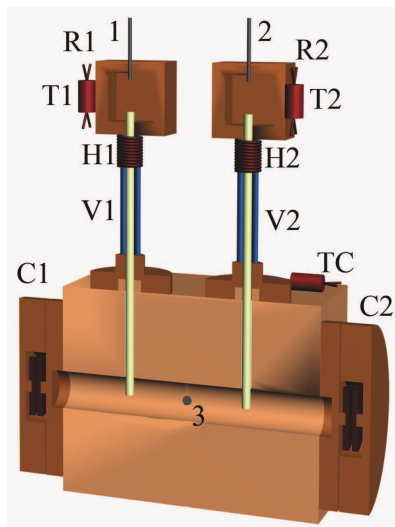}%
\caption{(Color online) Schematic rendition of the sample cell (not to scale), which consists of two Vycor rods, $V1$ and $V2$, reservoirs, $R1$ and $R2$, their heaters, $H1$ and $H2$, and a cylindrical space for the helium sample ($1.84$~cm$^3$) with a Straty-Adams pressure gauge \cite{Straty1969}, $C1$ and $C2$,  at each end of the solid helium region.  A chemical potential difference can be applied between the two reservoirs. $TC$, $T1$ and $T2$ are calibrated resistance thermometers for the helium sample cell and for the two liquid helium reservoirs, respectively. Filling capillaries 1 and 2 lead to reservoirs $R1$ and $R2$ and capillary 3 is thermally connected to the 1~K pot and still and enters from the side and provides direct access to the helium sample space for efficient initial filling of the cell. \label{cell}}
\end{figure}

To study mass flow through solid helium-4 an apparatus was designed and was previously described in some detail \cite{Ray2008a,Ray2009b,Ray2010a,Ray2010b,Vekhov2012} (see Fig.~\ref{cell}). A solid \4he sample is situated in series between two Vycor (porous glass with interconnected pores of diameter $\approx$ $7$~nm) rods with bulk liquid reservoirs on the top of each rod. The Vycor rods are 0.140 cm in diameter, 7.620 cm in length,
and the cylindrical surface of the Vycor external to the reservoirs and the cylindrical chamber that houses the solid helium
is sealed with a thin coating of Stycast 2850 FT epoxy.  This configuration allows for the application of a chemical potential difference, $\Delta\mu$, across the solid helium sample. The initial chemical potential difference, $\Delta\mu_0$, can be imposed either by the application of a pressure difference between the two reservoirs, e.g. by injection or withdrawal of atoms from one or both reservoirs\cite{Ray2008a,Ray2009b}), or by the utilization of the fountain effect by the application of a temperature difference, $\Delta T$, between the two superfluid-filled reservoirs\cite{Ray2010b,Ray2010c,Vekhov2012}.

To introduce a liquid helium sample into the cylindrical sample cell ($V$ = 1.84 cm$^3$) and ultimately reach the desired pressure one can use a combination of a side-entry direct-access capillary (labeled as $3$, Fig.~\ref{cell}) and lines $1$ and $2$ and condense helium gas (ordinary well helium with \emph{presumed} nominal $0.3$~ppm $^3$He impurity) to the horizontal cylindrical space between the two \emph{in situ} pressure gauges, $C1$ and $C2$, at a constant temperature of the sample cell.  The capillary lines that enter the cell have an inside diameter of 0.13 mm and a length of $\approx$ 1.52 m.

The growth of a solid helium sample is typically started from the superfluid, so the first stage of the solid growth procedure is similar to the creation of a liquid sample. After approaching the melting curve the direct line ($3$ in Fig.~\ref{cell}) is closed while helium injection through lines $1$ and $2$ is continued. To increase the pressure in the sample cell above the melting curve\cite{Ray2010c}, we find that the sample cell temperature has to be in the range $0.3 - 0.4$~K. We find that it is very hard to cause a sample to leave the melting curve when $T < 0.3$~K.  After the pressure has moved from the melting curve, further growth of solid helium can be accomplished with the temperature at selected values in the range  $0.1 \lesssim T \lesssim 0.4$~K. Usually the desired pressure for a solid sample is reached after several hours  of growth. After such growth we allow the system to stabilize for a few additional hours at $T < 0.4$~K. All of the samples studied here were fresh grown and not annealed at temperatures above the range of temperatures studied.

The pressure range accessible in the present experimental configuration has an upper limit that is imposed by the need to keep the liquid helium in the reservoirs in the superfluid phase between the melting curve and the $\lambda$-line on the $P-T$ phase diagram. There is a temperature gradient maintained along each Vycor rod from the lower solid sample temperature to the higher liquid helium reservoir temperature. This is to ensure that the helium in the reservoirs does not solidify. Based on these conditions, an approximate upper limit for the cell pressure for the measurements reported here (as measured by gauges $C1$ and $C2$) for the technique we use is  $\sim 26.1$~bar. Due to the fountain effect,
\begin{equation}
\Delta P_f = \int_{T_a}^{T_b} \rho s dT,
\label{Eq_fountain}
\end{equation} 
the pressures in the reservoirs, $P1$ and $P2$, are higher than the \emph{in situ} cell pressures measured by $C1$ and $C2$.  During the course of our measurements, $C1$ and $C2$ drifted slightly with a similar mbar span of drift also seen for $P1$ and $P2$. These small drifts resulted from changes in the liquid helium level in the main 4.2~K liquid helium bath and are often present for most samples studied in our apparatus. Their presence does not influence the conclusions that we reach.

Once we have a sample of \4he in the cell, we use the heater H1 (H2) to vary $T1$ ($T2$) to create chemical potential differences between the reservoirs and then measure the resulting changes\cite{Ray2010b} in the pressures $P1$ and $P2$ (and $C1$ and $C2$).
 We take
the derivative of $P1 - P2$ with respect to time,
\begin{equation}
F = \frac{d(P1-P2)}{dt},
\label{Eq_F}
\end{equation} 
 and assume it to be proportional to the flux of atoms that passes through the solid-filled cell to get from one reservoir to the other, although it is the case that during mass flux some atoms are added to the solid and increase its density\cite{Ray2009b}.  We study $F$ as a function of $T$, $P$ and $\Delta \mu$, the chemical potential difference between R1 and R2, where, as we have noted, $\Delta \mu = m_4[\int(dP/\rho) - \int(sdT)]$.  We report $\Delta \mu$ in units of J/g.  We will report our flux values in mbar/s, where a typical value of 0.1 mbar/s corresponds to a mass flux through the cell of $\approx$ $4.8 \times 10^{-8}$ g/s.
To utilize the fountain effect \cite{Ray2010b} to induce a flow of atoms it is necessary, of course, to have superfluid helium inside the  Vycor rods\cite{Beamish1983,Lie-zhao1986,Adams1987}.

\section{Mass flow through liquid helium}

A number of measurements have been carried out with cell pressures below 24 bar with superfluid in the sample cell to determine some of the characteristics of the apparatus including flux limitations imposed by the Vycor.  A discussion of these diagnostic-type measurements made with no solid in the cell is deferred to Appendix A.

\section{Mass flow through solid helium}
\label{solid helium}

\subsection{Determination of Appropriate Protocols}

\begin{figure}
\resizebox{3.5 in}{!}{
\includegraphics[width=1\linewidth,keepaspectratio]{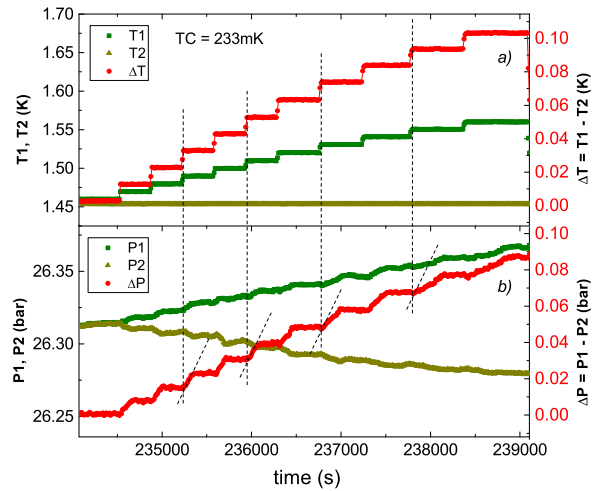}}%
\caption{ \label{fig:Limit-233-2} (color online)
Dependence of changes in pressures that accompany changes in $T1$ temperatures for a cell temperature, $TC =233$~mK. Changes in pressures $P1$ and $P2$ are seen, which result from step-wise increases of $T1$ in increments of 10~mK, with $T2$ constant. Short dashed tilted lines each with the same constant slope are guides to the eye to help reveal the presence of sequential changes in the slope of $\Delta P$.} 
\end{figure}

Once $\Delta T = |T1 - T2|$ is applied to create an initial finite $\Delta\mu_0$  one can document the kinetics of   $\Delta P = P1 - P2$, with $\Delta\mu$ decreasing and approaching zero with time, $t$.
An example of this for a solid \4he sample is shown in Fig.~\ref{fig:Limit-233-2} where the general behavior of $P1 - P2$ is shown for a sequence of steps in the reservoir $R1$ temperature, $T1$, and a fixed value of $T2$.  This sample, as was the case for all samples in this report, was grown from the melting curve by helium injection in the temperature range $0.3 < T < 0.4$~K.

\begin{figure}
\includegraphics[width=1\linewidth,keepaspectratio]{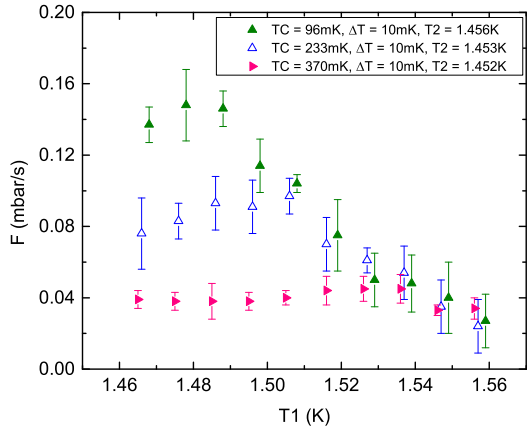}%
\caption{ \label{fig:FvsTR} (Color online) Dependence of the flow rate, $F$, resulting from 10 mK changes in the reservoir temperature, $T1$, for fixed $T2$ with the pressure of the solid in the cell $ 26.11 \pm 0.06$ bar (as measured by the gauges $C1$ and $C2$). Here $T1$ is the resulting temperature of reservoir R1 following each step of $\Delta T = 10$ mK, and $F$ is the resulting maximum value of the flux observed for each step in $T1$.  The data for $TC=233$~mK are obtained from the data for $TC=233$~mK shown in Fig.~\ref{fig:Limit-233-2}. 
}
\end{figure}

For these  measurements with solid \4he, primarily designed to confirm the acceptable range of $T1$ and $T2$ temperatures, the procedure was the following. With $T2$ stable, the temperature $T1$ is raised in a step-wise manner with incremental steps with the result that $\Delta T$ is increased in increments of  10~mK; Fig.~\ref{fig:Limit-233-2}(a) (green square data points). For each temperature step a change of the pressure difference, $P1-P2$, to a new stable state takes place (red circle data points, Fig.~\ref{fig:Limit-233-2}(b)).
One can see by inspection that the flow rate decreases when $T1$ increases above $\approx$ 1.49~K, presumably due to the reduced superfluid component in upper region of the Vycor rod (see Appendix A) adjacent to the reservoir, $H1$. Thus, to reliably measure flow rates through solid helium the temperature of the liquid helium reservoirs should not exceed $\approx 1.49$~K. Measurements of the sort shown in Fig.~\ref{fig:Limit-233-2} were taken at several cell temperatures, $TC$. The resulting \emph{maximum} flux values are shown in Fig.~\ref{fig:FvsTR}.
For the highest solid \4he sample temperature shown there is no evidence (within the noise of the data points) of a  flow rate change with an increase in $T1$ in the full range studied. This means that $F$ through a solid sample at high temperatures, where the flow rate is slow enough \cite{Ray2011a}, can be measured with no risk of limitation by flow through the Vycor rods up to at least $T1=1.54$~K. But, for uniformity we make all measurements under the same conditions (the same range of $T1$ and $T2$, $Ti \lesssim 1.49$~K for different solid helium temperatures), unless otherwise noted.

\subsection{Experimental Approach, Data and Characterizations}

\begin{figure}
\includegraphics[width=1\linewidth,keepaspectratio]{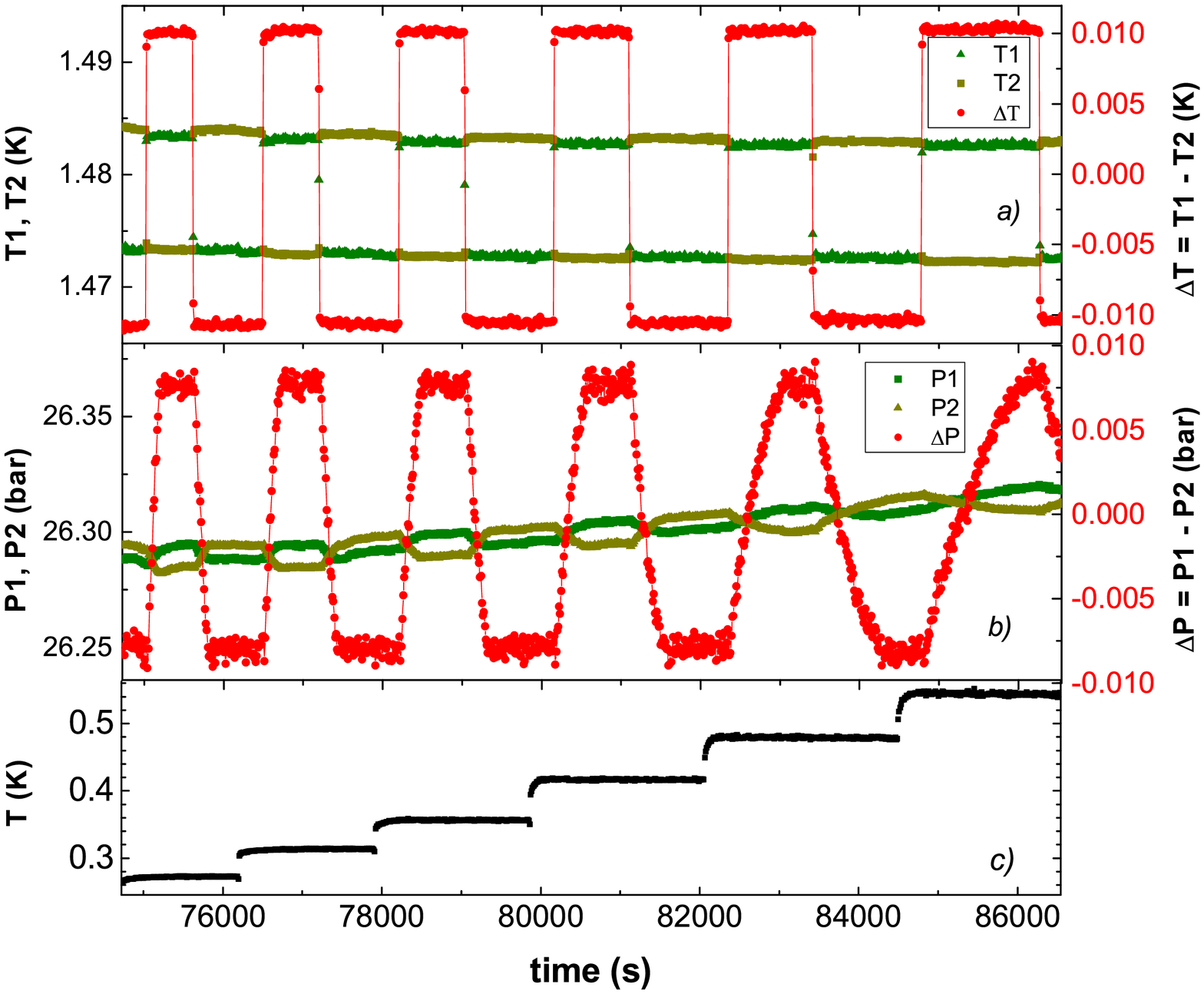}%
\caption{(Color online) Response of pressures $P1$ and $P2$ to the application of sequential $\Delta T$ reversals (see text) with $\delta T$ = 5 mK for a sequence of solid helium temperatures, here $0.25 \lesssim TC \lesssim 0.55$ K. Use of heaters, $H1$ and $H2$, results in changes in $T1$ and $T2$. The resulting changes in $P1$ and $P2$ are best seen as $P1-P2$, shown here (b, red circles). The small drift in $P1$ and $P2$ of the sort seen here is typical and variable and appears to have no significant influence on $P1-P2$.  \label{dPdTkinet}} 
\end{figure}

A number of flow measurements (without exceeding $Ti\approx 1.49$~K) were carried out for solid samples in the temperature range  $180 < T < 545$~mK.  
An example of data taken over a range of solid \4he temperatures for data from the same sample that was used for the data in the previous two figures is shown in Fig.~\ref{dPdTkinet}.
To obtain these data we utilize a technique modified from that described earlier. We simultaneously change both reservoir temperatures, but in opposite directions. To accomplish this a baseline reservoir temperature is first selected, $T_0$, with $T1 = T2 = T_0$. Then $T1$ is decreased by $\delta T$ while $T2$ is increased by the same interval; $\Delta T = T1 - T2 = -2\delta T$. After chemical potential equilibrium is reached (e.g., see  Fig.~\ref{dPdTkinet}), the values of $T1$ and $T2$ are interchanged
 so that $\Delta T = T1 - T2 = +2\delta T$. Each $Ti$, $Tj$ interchange results in a swing of the difference between the reservoir temperatures of $4\delta T$.
 With each switch in the value of $T1-T2$ there is a response of $P1-P2$. For a given value of the temperature difference between the two reservoirs this approach is expected to create a smaller perturbation on the solid that would our previous approach in which one $Ti$ was held fixed and the other changed, i.e. the density of the solid sample is not so much changed due to the so-called\cite{Soyler2009} ``syringe effect'' and it allows us to obtain larger $\Delta \mu_0$ values without exceeding the upper Vycor temperature at which a measurable flow limitation is encountered \cite{Vekhov2012}.

\begin{figure}
\includegraphics[width=1\linewidth,keepaspectratio]{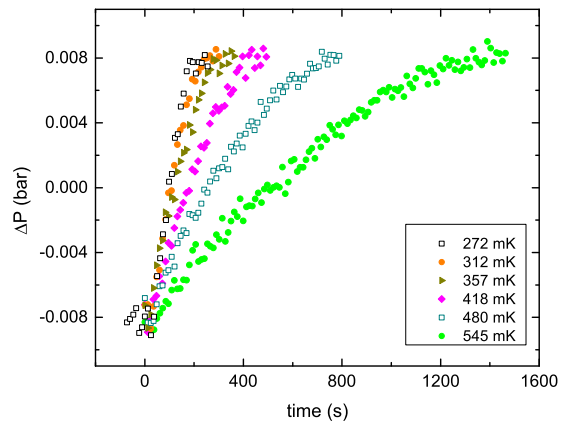}%
\caption{(Color online) Behavior of the pressure difference, $\Delta P$,  after applying a positive $\Delta T$ ($\delta T = 5$~mK) between the two liquid helium reservoirs at different solid helium temperatures.  These data are extracted from data sets like that shown in Fig.~\ref{dPdTkinet} with $P1-P2$ increasing.  The temperatures shown are the cell temperatures, $TC$. Similar behavior is present for $P1 - P2$ decreasing. \label{dPvstime-solid}} 
\end{figure}

Next, we compare the change in  $\Delta P$ $vs.$ time after application of a positive $\Delta T = T1-T2$ for constant $\Delta T$ values but for different solid helium temperatures, $TC$. In Fig.~\ref{dPvstime-solid} one can see a substantial qualitative influence of the temperature of the solid on the behavior of $\Delta P$ \emph{vs.} time for different solid \4he temperatures: the higher the \4he solid temperature the slower the relaxation. This agrees with earlier observations \cite{Ray2011a}, where above 100 mK the flux decreased with an increase in the solid helium temperature.


\begin{figure}
\includegraphics[width=1.0\linewidth,keepaspectratio]{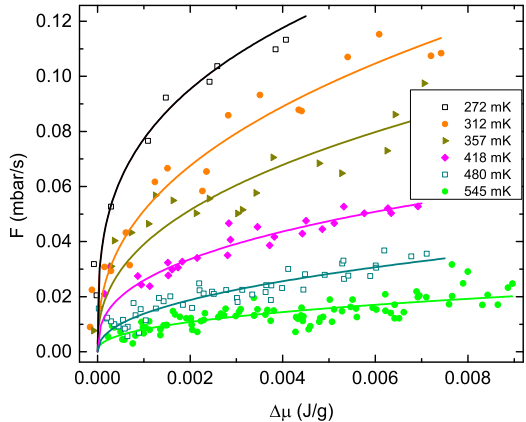}
\caption{(Color online) The flow rate, $F$, vs. $\Delta\mu$ after application of $\Delta T > 0$ between the two liquid helium reservoirs for different solid helium temperatures determined from the data shown in Fig.~\ref{dPvstime-solid}. Solid lines here are fits to the data by $F = A(\Delta\mu)^b$.
\label{Fvsdmu-solid22}} 
\end{figure}

The data shown in Fig.~\ref{dPvstime-solid} is taken from the $\Delta T > 0$  portion of the data sets of the type seen in
Fig.~\ref{dPdTkinet}.   Approximately similar data can be found for the case of $\Delta T < 0$. From such data we proceed to document how the flow rate, $F$, depends on $\Delta\mu$. To use the specific example of data already presented,  data of the sort shown in Fig.~\ref{dPvstime-solid} are averaged (three points for the data at $T$=272, 312 and 357~mK, nine points for the data at 418 and 480 mK, and twelve points for the 545 mK data)  and then differentiated according to Eq.~(\ref{Eq_F}) by use of a three-point algorithm and $\Delta\mu$ is calculated according to Eq.~(\ref{Eq_chemical}).  The result for $F$ \emph{vs.} $\Delta\mu$ is shown in Fig.~\ref{Fvsdmu-solid22} for different solid helium temperatures and for  $\delta T$ = 5 mK.

\begin{figure}
\includegraphics[width=1.0\linewidth,keepaspectratio]{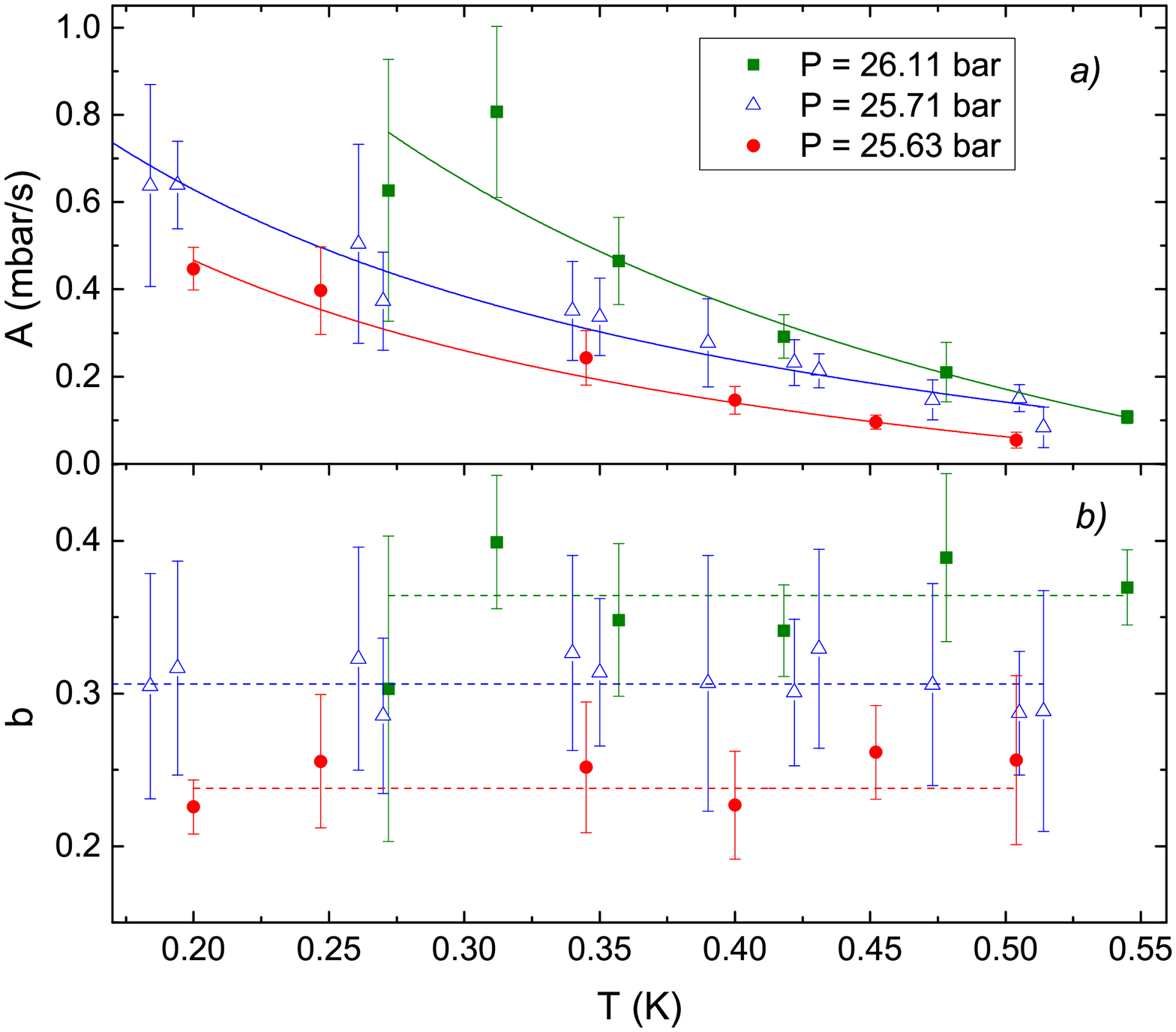}%
\caption{(Color online) Temperature dependence of the fit parameters $A$ and  $b$ for three solid \4he samples each with different cell pressure. Lines are guides to the eye.  \label{b2-pressure}} 
\end{figure}

 A power law is chosen as a good candidate to characterize these data\cite{Vekhov2012}.  A comparison of a power law \emph{vs.} an exponential    shows that the power law gives smaller residuals for most of the data sets (although for the lowest temperature data the two different fits produce roughly the same goodness of fit). Thus a power law (with two parameters as opposed to three parameters for the exponential) has been chosen to fit all the data sets,
\begin{equation}
F = A(\Delta\mu)^b,
\label{Eq_FvsDmu}
\end{equation} 
where $A$ and $b$ are fit parameters. As will be discussed later, this functional dependence (with $b < 0.5$) is consistent with the non-linear behavior expected for a Luttinger liquid.  The temperature dependence of these parameters for three samples at different pressures (as determined by the \emph{in situ} pressure gauges $C1$ and $C2$) is shown in Fig.~\ref{b2-pressure}. Parameter $A$ decreases monotonically with increasing temperature, while $b$ is independent of temperature within our error bars; both depend on pressure.



\begin{figure}
\includegraphics[width=1.0\linewidth,keepaspectratio]{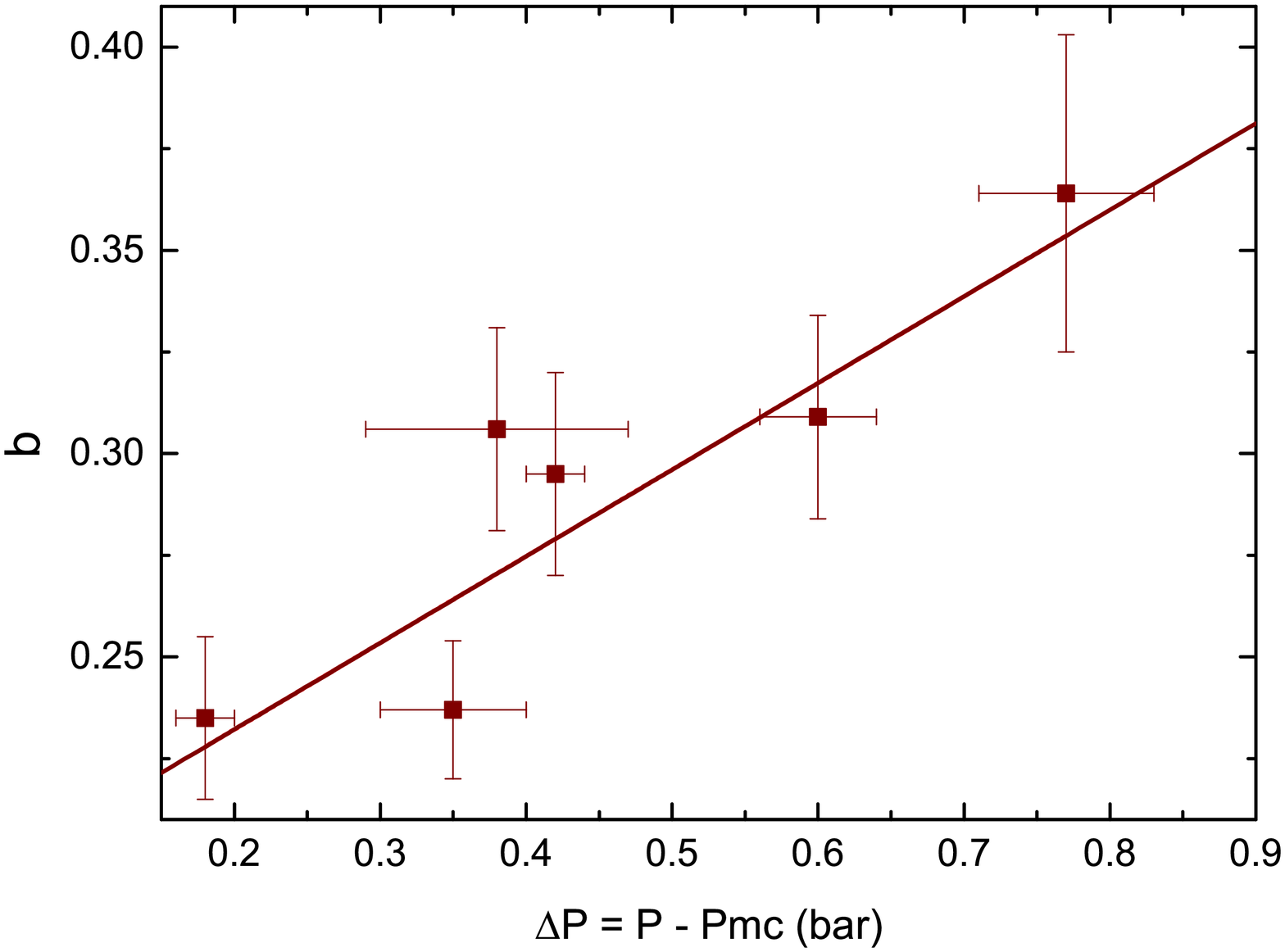}%
\caption{(Color online) Pressure dependence of the fit parameter $b$ determined for samples with six different solid \4he sample pressures; $Pmc = 25.34$ bar. The line is a linear fit (see text). \label{b-pressure}} 
\end{figure}

The dependence of the parameter $b$ on the distance of the pressure of the solid in the cell from the melting curve pressure, denoted as $\Delta P = P - P_{mc}$, (Fig.~\ref{b-pressure}) is  characterized reasonably by a linear fit, $b = \alpha + \beta(P-P_{mc})$, with $\alpha = 0.19 \pm 0.02, \beta = 0.21 \pm 0.07$.

\begin{figure}
\includegraphics[width=1.0\linewidth,keepaspectratio]{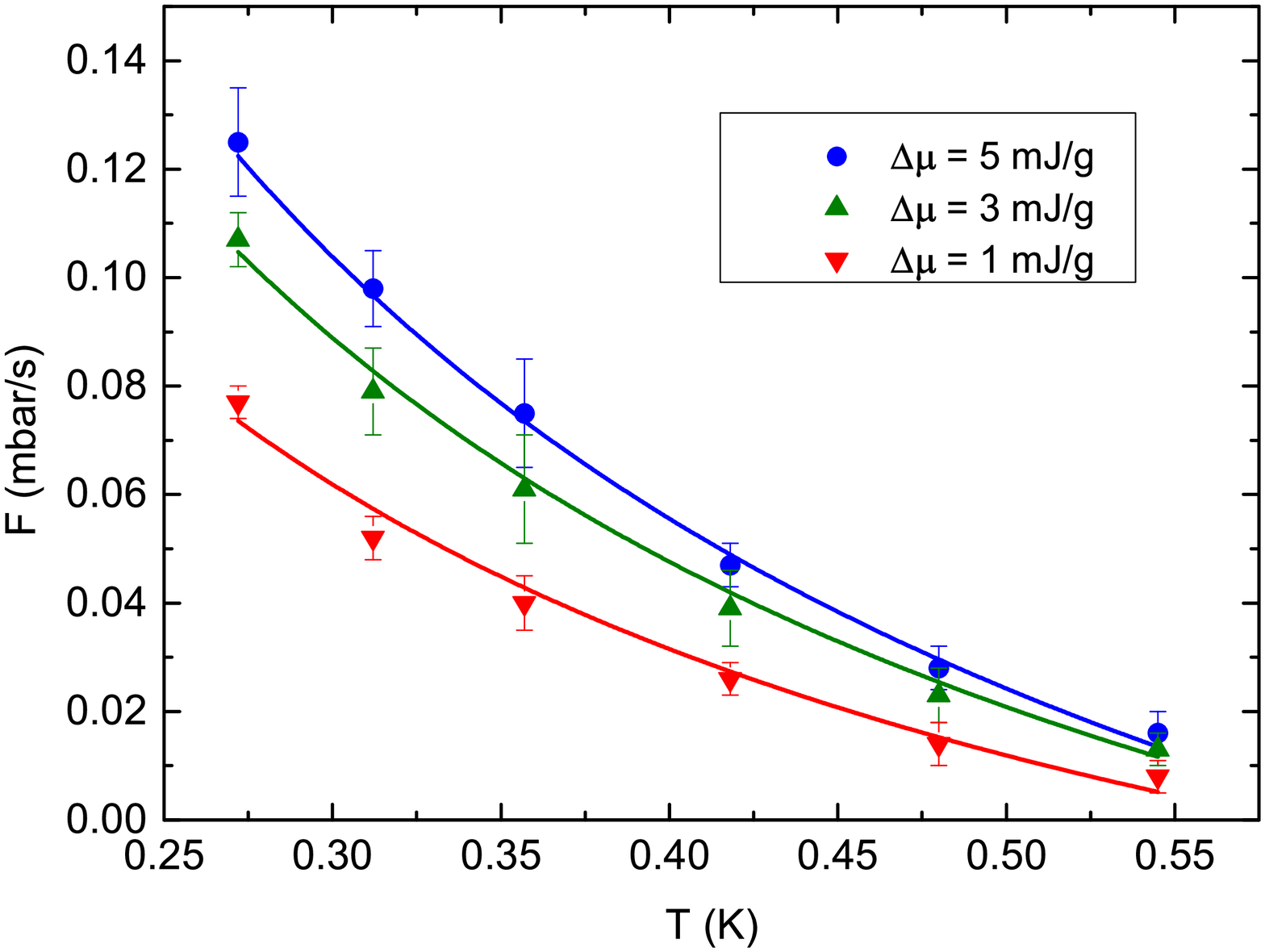}%
\caption{(Color online) Temperature dependence of the mass flow, $F$, through solid helium for different fixed $\Delta\mu$ values at a cell pressure of 26.11 $\pm 0.1$ bar. Solid lines are the result of fits to Eq.(\ref{Eq_FvsT}) with $B/A$ = 1.21 and $E = 117 \pm 2, 117 \pm 2$ and $112 \pm4$ for $\Delta\mu = 5, 3$ and $1$ mJ/g respectively.  \label{FvsT-solid}} 
\end{figure}

The temperature dependence of $F$ is plotted in Fig.~\ref{FvsT-solid}  for three different \emph{fixed} $\Delta\mu$ values and fixed pressure for a single sample. The flow extrapolates to values too low to be measurable at a characteristic temperature, $T_h \sim 630$ mK.	A fit of these data sets and others to the functional form
\begin{equation}
F = A - B\exp(-E/T),
\label{Eq_FvsT}
\end{equation} 
where $A, B$ and $E$ are fit parameters, results in reasonable fits for $B/A = 1.21 \pm 0.06$, with an average value $E = 0.12 \pm 0.02$K for the pressure range we have studied. The dependence of $A$ on pressure seen in
Fig.~\ref{b2-pressure} suggests that $E$ has a pressure dependence and we will explore this further in future work.
We find that $F = F_0^*[1 - 1.21\exp(-E/T)]$ can be applied to individual data sets, with the interpretation that $F_0^*$ should in each such case be proportional to the number of conducting pathways between the Vycor rods.  We have determined that the temperature dependence is a universal function of temperature with  $F^*_0$ dependent on the particular sample, its history\cite{Vekhov2013} and the $E(P)$ dependence mentioned above.

The functional form of this temperature dependence used here, Eq.~(\ref{Eq_FvsT}), is an improvement over that we used earlier to characterize data of this type\cite{Vekhov2012}, $F = - z\ln(T/\tau)$.  The present dependence is better motivated physically and for the same number of parameters (two) it results in a better goodness of fit.  The present functional form suggests the possibility that
a thermally activated process exists that degrades the flux with increasing efficiency according to $\sim \exp(-E/T)$. For example, thermally activated jogs or kinks\cite{Aleinikava.2012} (roughness) on dislocation cores would introduce disorder and phase slips would result and reduce the flux.


\section{Discussion}

To understand the nature of the mass flow through solid helium it is useful to consider different possible scenarios.  Generally speaking, the conducting pathways could be three, two or one-dimensional. For the case of three dimensional paths bulk transport might take place through the entire volume of the solid. One can also consider bulk liquid channels, which can form between solid helium and  sample cell walls or along contact between two grain boundaries of helium polycrystals and the sample cell wall\cite{Sasaki2008}. For the case of 2D paths one can consider grain boundaries which have been predicted to be superfluid \cite{Pollet2007}. Finally, to discuss one-dimensional conducting paths we consider the predicted superfluid cores of screw \cite{Boninsegni2007} or edge dislocations \cite{Soyler2009}.  Three-dimensional liquid channels could also become one-dimensional channels if they became narrow enough.  Such behavior would involve a transition from three-dimensional to one-dimensional behavior as a function of pressure.

It is known that for a bulk superfluid the mass flow, $F$, does not depend on the chemical potential difference applied, $\Delta\mu$,  in a readily measurable way until a substantial flux is present. The rapidly rising dissipative behavior with increasing flux becomes measurable and results in what is often referred to as a ``critical velocity".
This approach to the dissipative regime in a superfluid system has at times been explored by study of the flow velocity associated with  pressure gradients of various sorts\cite{Kidder1962,Flint1974,Chan1980} under quasi-isothermal conditions.  As we have pointed out previously\cite{Vekhov2012} interchange of the axes on figures like Figure \ref{Fvsdmu-solid22} provide a representation that is reminiscent of such studies.

An apparent critical flow (or behavior close to that) was observed for solid helium on the melting curve for 3D paths along grain boundaries for helium polycrystals in contact  with the sample cell wall \cite{Sasaki2008}.  Such paths are predicted\cite{Sasaki2008} to have a cross sectional area that depends rather strongly on pressure.  In our typical pressure range of $0.2 \lesssim P - P_{mc} \lesssim 0.8$ bar for the solid in the cell, the predicted cross sectional area, $\Lambda$ of such paths is $760 \gtrsim \Lambda \gtrsim 50$ nm$^2$, which would correspond to cylindrical tubes with effective diameter, $D$, $31 \gtrsim D \gtrsim 8$ nm.   We are not aware of superfluid flow measurements in channels in this diameter range, but torsional oscillator measurements have been used to study the superfluid density in channels with diameters in the range $4.7 \gtrsim D \gtrsim 1.5$ nm\cite{Ikegami2007,Taniguchi2010}. These studies of the temperature dependence of the superfluid density have been interpreted to show a transition to 1D-like behavior only in the vicinity of channel diameters of $\approx$ 1.8 nm.  It is also the case that the flux values we measure as a function of temperature (for nominal 300 ppb \3he impurity) drop very abruptly near 75-80 mK (and typically partially recover for lower temperatures)\cite{Ray2010c,Ray2011a}, a behavior not seen for flow in bulk-like liquid-filled channels\cite{Sasaki2006}, or for flow in Vycor\cite{Kiewiet1975}.  The temperature dependence of the flux we observe is different  and would appear to rule out small-diameter macroscopic 3D paths as candidates for the mass flow observed in our present experiments.

It is worth noting here that at the pressures of our experiments, $\lesssim$ 26.1 bar,  this temperature of 75-80 mK is not far from the predicted phase separation temperature, $T_P^S$ of solid mixtures of 300 ppb concentration, $\chi$.  At 26 bar we use
Eq.~(\ref{Eq_PS})
to calculate\cite{Ganshin2000,Edwards1989} this temperature to be $T_P^S$ = 62 mK,
\begin{equation}
T_p^s = [(0.80)(1 - 2\chi) + 0.14]/\ln(1/\chi -1),
\label{Eq_PS}
\end{equation}
but  Eq.~(\ref{Eq_PS}) properly applies to solid-solid phase separation and at our pressures, if the \3he separates  we expect that it will be a liquid.  When we take this into account we find that the temperature of the abrupt drop in the flux on cooling remains above the predicted homogeneous phase separation temperature\cite{Vekhov2013}.
Our on-going work involves other concentrations so we can explore more fully, among other things, how this abrupt drop in flux depends on concentration. Our initial observations indicate that as the concentration increases so does the temperature at which the abrupt drop in flux takes place\cite{Vekhov2013}, but it remains above the predicted bulk phase separation temperature.  There is limited experimental data in the literature on solid phase separation in our experimental regime\cite{Kim2011}.
 Work by Edwards et al.\cite{Edwards.1962} indicates deviations from $T^3$ behavior in the specific heat for $T > T_p^s$ that  suggests local \3he concentration fluctuations\cite{Antsygina.2005}, which may be relevant.

With regard to 2D paths it was shown that a decrease in the superfluid film thickness leads to dissipative behavior that is a rapidly increasing function of decreasing film thickness, with dissipation becoming readily measurable\cite{Telschow1976} once the thickness falls below a nominal value of $\sim$ 12 atomic layers. The temperature dependence of the superfluid density and dissipation for 2D helium films obeys the Kosterlitz-Thouless (KT) prediction\cite{Bishop1980}. Our flux measurements do not have the KT temperature dependence, which apparently rules out two dimensional liquid-like flows. It also apparently rules out the sorts of behavior seen for filled channels with diameters near 5 nm, where a temperature dependence of the superfluid density weaker than that given by Kosterlitz-Thouless has been observed\cite{Ikegami2007}.

Solid helium typically has a rather large (sample dependent) number of dislocations. Some of the first evidence for this in solid helium was obtained from sound velocity experiments\cite{Wanner1976}. A variety of experimental and theoretical studies of dislocations in solid helium have revealed a number of their properties. One of the first theoretical predictions of the possibility of superfluidity along the cores of edge dislocations was that proposed by Shevchenko\cite{Shevchenko1988}. Recent experimental thermodynamic studies of solid helium (precise pressure measurements)\cite{Grigorev2007a} observed an additional pressure $P\sim T^2$ contribution, which at $T<0.35$~K exceeds the phonon contribution. This $P\sim T^2$ contribution could be due to either a glass state or due to dislocations. Supposing dislocations, the dislocation density required to describe this additional pressure was estimated \cite{Lisunov2012} to be $N\sim 10^{12}$~cm$^{-2}$; recent work suggests a much smaller number\cite{Fefferman2014}.

Recent quantum Monte-Carlo (QMC) simulations have shown that liquid helium in a confined 1D geometry can be an example of a Luttinger liquid\cite{DelMaestro2010,DelMaestro2011}. The Luttinger liquid theory was developed many years ago for Fermi-systems\cite{Luttinger1963, Mattis1965}.  Based on the structure of long-distance correlations of a spinless fluid, Haldane \cite{Haldane} showed the essential similarity of one-dimensional Bose and Fermi fluids. Boninsegni \textit{et al.} \cite{Boninsegni2007}, also using QMC simulations, have predicted that superfluid cores of screw dislocations in solid helium could behave as a Bosonic Luttinger liquid.  For Luttinger-like behavior the chemical potential difference, $\Delta\mu$, will cause some kind of flow (e.g., electrical current, spin or mass flow) for which the current, $I$, is a non-linear function of the applied chemical potential difference $I\sim (\Delta\mu)^p$, where the exponent $p$ is related to the  Luttinger parameter, $g$.

\begin{figure}
\includegraphics[width=1.0\linewidth,keepaspectratio]{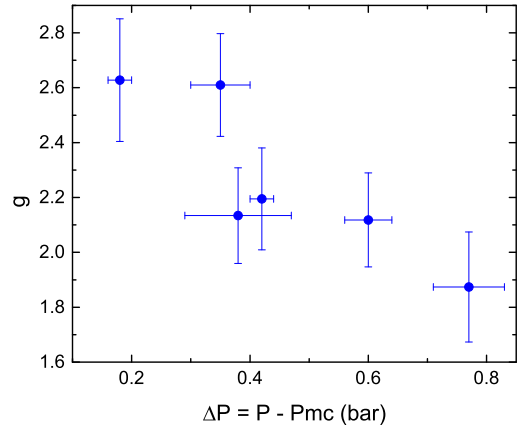}%
\caption{(Color online) Pressure dependence of the Luttinger parameter, $g$, in terms of the distance from the melting curve
presuming that $g = [(1/p) + 1]/2$, with $p = b$; again here, $Pmc = 25.34$ bar. \label{Luttinger}} 
\end{figure}

For our case, $\Delta\mu$ applied between the tops of the Vycor rods causes a mass flow of \4he atoms, $F$, through the solid helium with the result that  $F=A(\Delta\mu)^b$, where $b$ is temperature independent  but depends on pressure (see Section~\ref{solid helium}). We have taken care to ensure that the flow limitation is not due to the liquid in the Vycor (see Appendix A), but instead is due to the presence of the solid in our cell.  In the  Luttinger liquid model in the quantum regime where the impedance is caused by impurities, the exponent is expected to be given by  $p=1/(2g-1)$, where the Luttinger paramter, $g$, is independent of temperature \cite{Svistunov2012}. Assuming this relationship, the Luttinger parameter is determined from our measurements of $b$ to be as shown in Fig.~\ref{Luttinger}.
The Luttinger parameter $g$ is seen to decrease with increasing pressure as would be expected as the system moves  toward an insulating state.  Based on the analysis above one can conclude that if mass flow through solid helium is provided by 1D defects, e.g. superfluid cores of edge dislocations, a Luttinger liquid model seems relevant.
For a single conduction channel with Luttinger liquid behavior, one expects such behavior for $k_BT/\hbar << J$, where $J$ is the flux in atoms/s.  For our work, e.g. at $T \sim 0.2$~K, with $\Delta \mu \approx$ 0.01 J/g, we have a flux of $J \sim 7 \times 10^{15}$ atoms/s.   $T \sim 0.2$~K results in $k_BT/\hbar = 2.6 \times 10^{10}$.  This indicates that for Luttinger liquid behavior to be relevant to our results, the effective number of conducting channels that carry flux, $N$, should be $\lesssim 2.8 \times 10^5$, a number not unlike a density reported recently in dislocation studies\cite{Fefferman2014}.

To understand the temperature dependence of the mass flow, Fig.~\ref{FvsT-solid}, we can consider two different possibilities: (1) The conductivity of each of the presumed conducting paths is temperature dependent or (2) the number of conducting paths present is temperature dependent; or both may be present. With regard to the first mechanism, $\rho_S$ for a Luttinger liquid is predicted to depend on temperature\cite{Kulchytskyy2013}. Speaking microscopically in this context, increasing the temperature should also increase the density of jogs and or kinks along a dislocation core. This will increase the disorder. Indeed, the functional dependence found, Eq. (5), is consistent with this picture\cite{Vekhov2013}.  In this context, it is perhaps of interest to note that the temperature dependence of the superfluid density deduced from torsional oscillator measurements for helium in microscopic channels\cite{Ikegami2007} has an interesting transition as a function of channel diameter. In particular, for 1.8 nm case the temperature dependence of the superfluid density\cite{Ikegami2007} can be reasonably well fit by Eq.~(\ref{Eq_FvsT}), with $E \sim 0.4$K.

Concerning the second mechanism, if we presume that superflux along edge dislocations is the cause of the flow, we need to consider the temperature dependence of the mobility of edge dislocations  first discussed theoretically by Granato and Lucke\cite{Granato1956} and confirmed in shear modulus experiments\cite{Paalanen1981} in solid $^4$He. A rising mobility of edge dislocations with increasing temperature decreases the number of their intersections, which in turn decreases the percolation of the whole dislocation net and can be expected to reduce the observed mass flow.

We should note here that typically if the solid \4he is warmed to $T > 630$ mK, on subsequent cooling no flux recovery or recovery to smaller flux values takes place unless atoms are subsequently added or withdrawn from the sample cell with a corresponding change in the cell pressure. One can perhaps assume that the addition or withdrawal of helium (density of the solid helium sample increasing or decreasing) leads to a redistribution of dislocations in the solid and the formation new intersections which recover the percolation and thus recover the flow through the solid.

An alternative mechanism to consider  is plastic flow. There are three main mechanisms of plastic flow or plastic deformation: (1) Diffusion; the diffusion mechanism is realized by moving vacancies and is characteristic of high temperatures. This means that rising temperature should lead to larger flow rates. We can exclude the diffusion mechanism because of our $F(T)$ dependence (see Fig.~\ref{FvsT-solid}). (2) Dislocations: this is due to the \emph{gliding} of dislocations and is not thermally activated. 
Gliding of dislocations should not stop as the temperature is increased. The gliding could stop during cooling due to dislocation pinning. We believe that we can exclude this dislocation mechanism because our flux ceases above 630 mK.  (3) Gliding along grain boundaries: this is more complicated because in solid helium it seems that grain boundaries can be superfluid \cite{Pollet2007}. Looking at our $F(T)$ dependence (and the absence of flow at high temperatures) the only gliding present could be gliding along superfluid grain boundaries (gliding along normal grain boundaries cannot be suppressed by temperature). But, we doubt that the one-dimensional-like dependence of flux on chemical potential we have observed would be present for gliding.
We also note that plastic deformation has a threshold strength, i.e. in the case of purely plastic flow we should see some residual $\Delta \mu$ without its relaxation to zero.

We believe that the evidence points most strongly to superfluid-like transport along dislocation cores as the likely cause of the flux that we observe.

\section{conclusions}

We have studied the flux of helium through a solid-filled cell as a function of temperature with a focus on temperatures above 180 mK.  We find that the flux is a non-linear function of the applied chemical potential. This is reminiscent of the behavior of a Luttinger liquid, which causes us to believe that whatever carries the flux through the solid-filled cell behaves like a bosonic Luttinger liquid.  The non-linear exponent is a function of pressure and the deduced Luttinger parameter decreases with an increase in pressure.  The flux at constant chemical potential decreases as a function of increasing temperature in a manner that suggests that thermally activated disorder is present.  A candidate for the flux conduit that is consistent with these characteristics is the cores of edge dislocations in the solid.



\section{Acknowledgment}

We thank M.W. Ray for his previous work on the apparatus and helpful comments and B. Svistunov and W. Mullin for a number of stimulating discussions. This work was supported by NSF DMR 12-05217, DMR 08-55954, and to a limited extent by DMR 07-57701, and also by Research Trust Funds administered by the University of Massachusetts Amherst.

\section{Appendices}

In Appendix A we discuss the characteristics of the apparatus when only superfluid \4he (no solid) is present.  We then go on in Appendix B to briefly discuss the effect of temperature excursions above $T_h$.

\subsection{Mass flow through liquid helium-filled cell}
\label{liquid helium}

  In previous work with this technique  a flow limitation introduced by the Vycor rods was seen when the liquid helium reservoirs had too high a temperature\cite{Ray2011a,Vekhov2012}, even when below the $\lambda$-temperature for Vycor. Thus, it is necessary to re-establish for the present work what the highest $T1$ and $T2$ values are that we can use to avoid this limitation and to study the flow through solid helium without a significant influence due to the Vycor rods.  And, it is important to determine if there are limitations on the response rate of the measured pressures and temperatures as a result of the application of temperature changes due to the heaters that induce the flow of helium to and from the reservoirs.

\begin{figure}
\resizebox{3.5 in}{!}{
\includegraphics{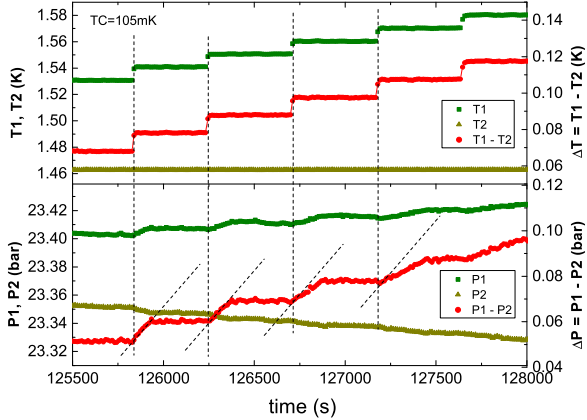}}
\caption{ (color online)  
 Response of pressures $P1$ and $P2$  to the application of several steps in $\Delta T$ for the case of liquid helium in the experimental cell at 23.2 bar and  a cell temperature of $T$ = 105 mK.  Use of heater H1 with  H2 fixed,
results in changes to $T1$ ($\approx$ 10 mK each) with $T2$ fixed.  The resulting $\Delta T$ and changes in $P1$ and $P2$ are shown and
best seen as $\Delta P = P1-P2$. Note that when the reservoir temperatures exceed $\approx$ 1.50 K the response of $P1$ and $P2$ slows as shown by the decrease in $d\Delta p/dt$ above 1.5 K (lines of equal constant slope are shown as guides to the eye); i.e., the flux is limited by the Vycor.
\label{fig:exampleL} }  
\end{figure}

Once $\Delta T = |T1 - T2|$ is applied to create an initial finite $\Delta\mu_0$  one can document the kinetics of   $\Delta P = P1 - P2$, with $\Delta\mu$ decreasing and approaching zero with time, $t$, as was done for the situation with solid \4he in the sample cell. This behavior, $\Delta P(t)$,  is shown in Fig.~\ref{fig:exampleL} for a sequence of  $T1$ increases of 10 mK applied for the case when the sample cell is filled with liquid \4he with $T2$ fixed at 1.462 K.  For this case it is apparent that when the reservoir temperature $T1$ exceeds $\approx$ 1.50~K  the response of the pressures $P1$ and $P2$ changes and becomes limited, and the limitation increases with increasing $T1$.  Thus, the apparatus places an upper limit on the flow rates when the reservoir temperature exceeds 1.50~K. This is likely the result of the limitation on $\rho_s$ in the Vycor at these elevated reservoir temperatures.

\begin{figure}
\resizebox{3.5 in}{!}{
\includegraphics{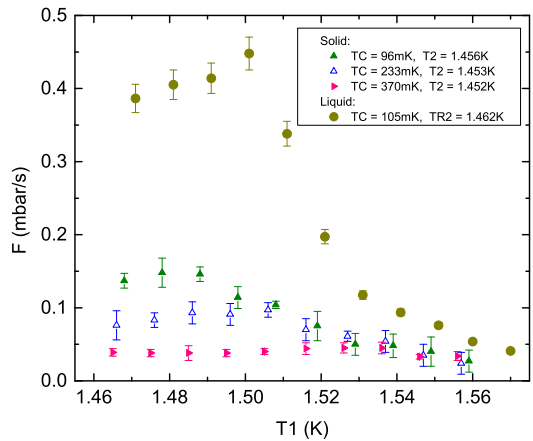}}
\caption{ \label{fig:liquid-flux2} (color online)
Maximum flux (calculated from the data shown in the previous figure) following the application of a temperature step $\Delta T \approx 10$ mK for the case of liquid helium in the experimental cell at 23.2 bar  for $T$ = 105 mK.   Also shown for comparison are the data for the case of solid helium in the cell (from Fig.~\ref{fig:FvsTR}).  Use of heater H1 or  H2
results in changes to $T1$ and $T2$. Shown here is the flux as a function of the reservoir temperature following the step in temperature.  In each case $Ti$ is changed with $Tj$ fixed at 1.462 K.  The time resolution of our data limits the flux to about 0.4 mbar/sec. Values for the solid-limited flux fall well below the values achieved when only superfluid \4he is in the cell.} 
\end{figure}

The use of equation (3)
gives us a measure of the flow rate of \4he through the sample cell expressed in the units of mbar/s. With \emph{liquid} helium in the apparatus, flows as high as $\approx 0.4 $ mbar/s are observed.  But, as shown in Fig.~\ref{fig:liquid-flux2} the Vycor does impose a limitation on the flux and this depends on which specific reservoir heater is employed.  Typically we ensure that the reservoir temperatures are below 1.49~K during our measurements of the flux with solid in the cell.  Flux values with solid helium in the experimental cell typically fall below 0.15 mbar/s and we are thus confident that the flux measurements we report are dominated by limitations imposed by the solid-filled cell, not by the conductivity of the Vycor.

\begin{figure}
\resizebox{3.5 in}{!}{
\includegraphics{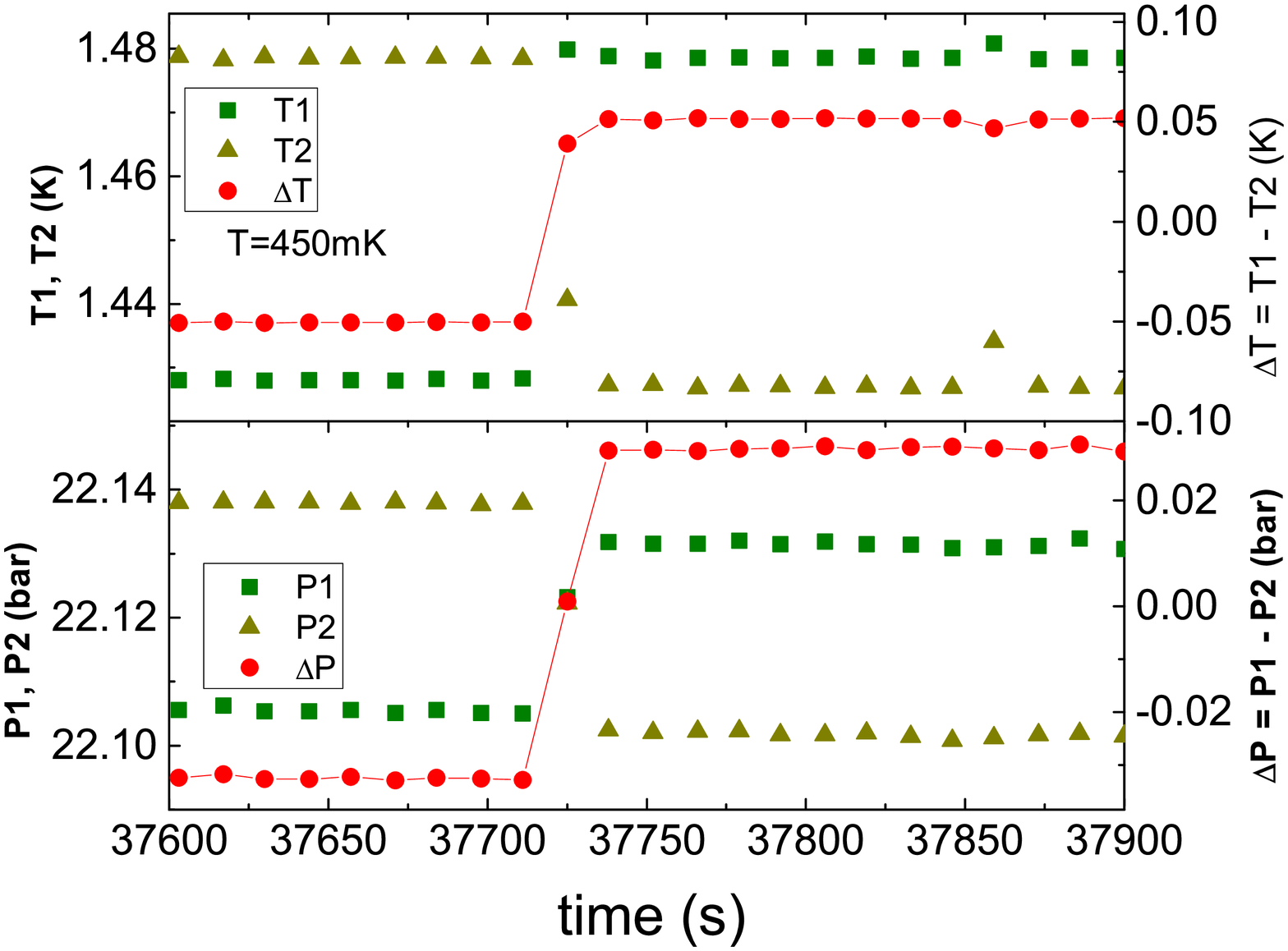}}
\caption{ \label{fig:liquid-1} (color online)
Response of pressures $P1$ and $P2$ to the application  of a temperature step $\Delta T$ for the case of liquid helium in the experimental cell at 22.0 bar and  $T$ = 450 mK.   
 Use of heater H1 and  H2,
results in changes to $T1$ and $T2$.  The resulting $\Delta T$ and changes in $P1$ and $P2$ are shown as is $\Delta P = P1-P2$. Lines through the data for $\Delta T$ and $\Delta P$ on this and the next figure are guides to the eye.} 
\end{figure}

\begin{figure}
\resizebox{3.5 in}{!}{
\includegraphics{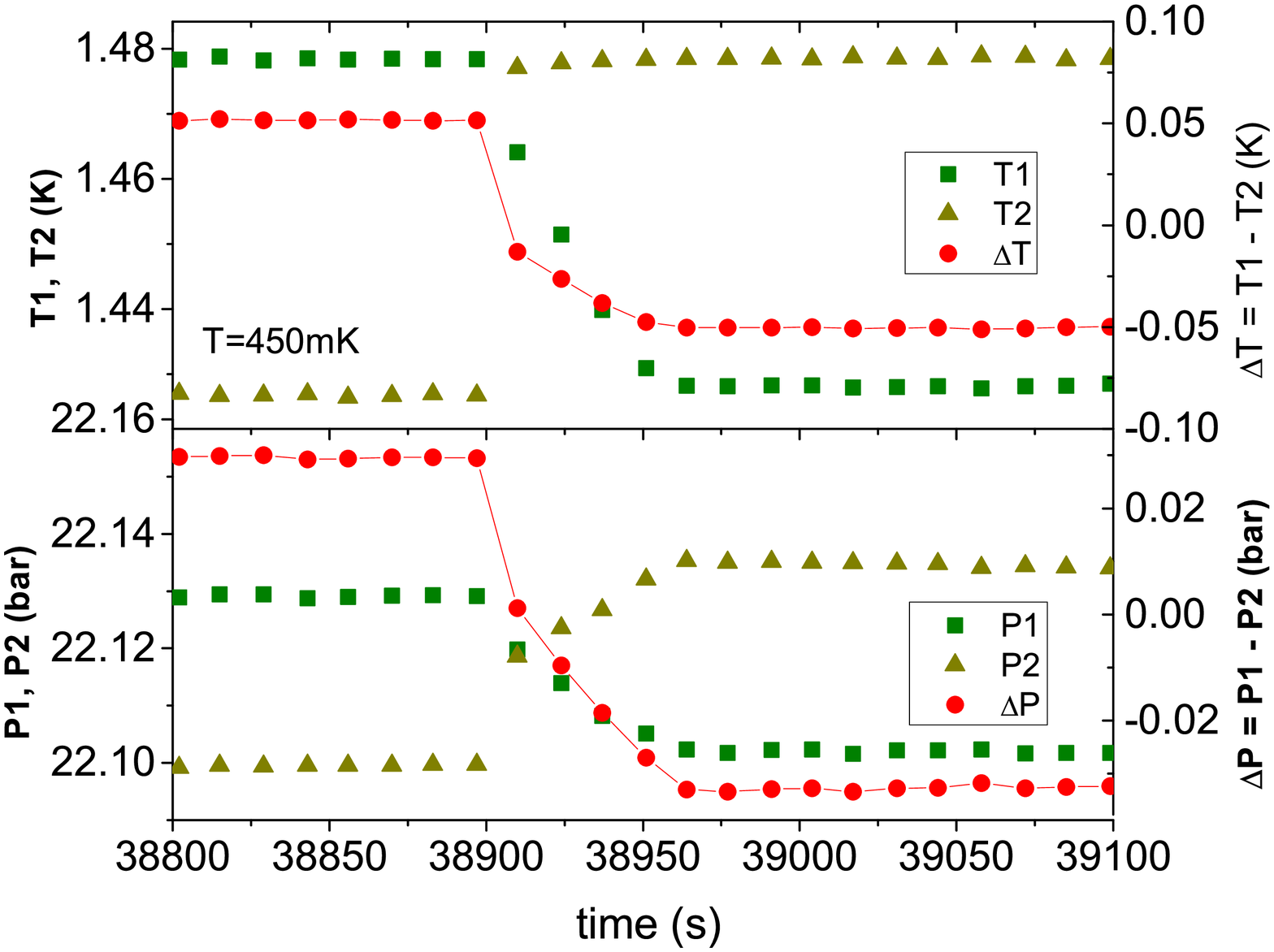}}
\caption{ \label{fig:liquid-2} (color online)
Response of pressures $P1$ and $P2$ to the application  of a temperature step $\Delta T$ for the case of liquid helium in the experimental cell at 22.0 bar and  $T$ = 450 mK.
Use of heater H1 and  H2, in this case,
results in changes to $T1$ and $T2$, which here are the reverse of those shown in the previous figure.  The resulting $\Delta T$ and changes in $P1$ and $P2$ are shown as is $\Delta P = P1-P2$. }  
\end{figure}

In addition to the effect of the reservoir temperatures on the flow, it is important to explore what the influence of the rate of change of the temperature might have on the response of the pressures.  To study this, we use a technique that was described more fully earlier in this manuscript when we used it for our measurements with solid in the cell.   But, in short, rather than increase one $Ti$ while keeping the other fixed, we increase $Ti$ and at the same time decrease $Tj$ by the same amount.  For reservoir temperatures that are well below 1.50 K, as shown in Fig.~\ref{fig:liquid-1} the responses of $P1$ and $P2$ to an increase in $T1$ and a decrease in $T2$ is prompt (within the resolution of our data collection rate). But as shown in  Fig.~\ref{fig:liquid-2} the response of $P1$ and $P2$ to a increase in $T2$ and a decrease in $T1$ is not as prompt. The reservoir(s) heat rapidly, but R1 cools a bit more slowly apparently due to a  conduction path to the refrigeration that has slightly more thermal impedance.

\subsection{Effect of temperature trajectories above $T_h$}

When the temperature of solid helium rises above $T=T_h$ the flow ceases and after cooling\cite{Ray2009b} does not recover to the original flux (or more often does not recover at all) unless the solid is manipulated by the addition or removal of atoms from the cell. After decreasing the temperature  to $\sim$ $100-300$~mK flow can typically be recovered by (1) helium withdrawal through the  two Vycor rods which leads to a pressure decrease in the cell; or (2) helium addition through the Vycor rods which leads to pressure increase in the cell. Presumably these pressure changes alter the disorder present in the solid.  The mechanism for this density change in the solid helium was proposed in Ref.~\cite{Soyler2009}, and was termed isochoric compressibility (or the ``syringe effect"), and is based on the so-called ``superclimb'' of edge dislocations. A separate publication will be devoted to an experimental study of this so-called isochoric compressibility.

\bibliography{ref3}

\end{document}